\documentclass[twocolumn]{aastex631}

\usepackage{xcolor}

\accepted{in ApJL \today}

\shorttitle{[CII] rotating disk in a low-mass dusty galaxy}
\shortauthors{Pope et al.}

\graphicspath{{./}{figures/}}

\begin{document}

\title{ALMA reveals a stable rotating gas disk in a paradoxical low-mass, ultra-dusty galaxy at $z=4.274$}

\author[0000-0001-8592-2706]{Alexandra Pope}
\affiliation{Department of Astronomy, University of Massachusetts, Amherst, MA 01003, USA}

\author[0000-0002-6149-8178]{Jed McKinney}
\affiliation{Department of Astronomy, University of Massachusetts, Amherst, MA 01003, USA}

\author[0000-0001-9394-6732]{Patrick Kamieneski}
\affiliation{Department of Astronomy, University of Massachusetts, Amherst, MA 01003, USA}

\author[0000-0003-4569-2285]{Andrew Battisti}
\affiliation{Research School of Astronomy and Astrophysics, Australian National University, Cotter Road, Weston Creek, ACT 2611, Australia}

\author[0000-0002-6590-3994]{Itziar Aretxaga}
\affiliation{Instituto Nacional de Astrof\'isica, \'Optica y Electr\'onica, Luis Enrique Erro 1, Tonantzintla CP 72840, Puebla, M\'exico}

\author[0000-0003-2680-005X]{Gabriel Brammer}
\affiliation{Cosmic Dawn Center (DAWN)}
\affiliation{Niels Bohr Institute, University of Copenhagen, Jagtvej 128, 2200 Copenhagen N, Denmark}

\author[0000-0001-9065-3926]{Jose M.~Diego}
\affiliation{Instituto de F\'isica de Cantabria (CSIC-UC). Avda. Los Castros s/n. 39005 Santander, Spain}

\author{David H.~Hughes}
\affiliation{Instituto Nacional de Astrof\'isica, \'Optica y Electr\'onica, Luis Enrique Erro 1, Tonantzintla CP 72840, Puebla, M\'exico}

\author{Erica Keller}
\affiliation{National Radio Astronomy Observatory, 520 Edgemont Road, Charlottesville, VA 22903, USA}

\author{Danilo Marchesini}
\affiliation{Department of Physics and Astronomy, Tufts University, Medford, MA 02155}

\author[0000-0003-2722-6600]{Andrew Mizener}
\affiliation{Department of Astronomy, University of Massachusetts, Amherst, MA 01003, USA}

\author[0000-0003-4229-381X]{Alfredo Monta\~na}
\affiliation{Instituto Nacional de Astrof\'isica, \'Optica y Electr\'onica, Luis Enrique Erro 1, Tonantzintla CP 72840, Puebla, M\'exico}

\author[0000-0001-7089-7325]{Eric Murphy}
\affiliation{National Radio Astronomy Observatory, 520 Edgemont Road, Charlottesville, VA 22903, USA}

\author[0000-0001-7160-3632]{Katherine E. Whitaker}
\affil{Department of Astronomy, University of Massachusetts, Amherst, MA 01003, USA}
\affil{Cosmic Dawn Center (DAWN)}

\author{Grant Wilson}
\affiliation{Department of Astronomy, University of Massachusetts, Amherst, MA 01003, USA}

\author{Min Yun}
\affiliation{Department of Astronomy, University of Massachusetts, Amherst, MA 01003, USA}

\begin{abstract}
We report ALMA detections of [CII] and dust continuum in Az9, a multiply-imaged galaxy behind the Frontier Field cluster MACSJ0717.5+3745. The bright [CII] emission line provides a spectroscopic redshift of $z=4.274$. This strongly lensed ($\mu=7\pm1$) galaxy has an intrinsic stellar mass of only $2\times10^{9}M_{\odot}$ and a total star formation rate of $26\,\rm{M_{\odot}/yr}$ ($\sim80\%$ of which is dust obscured). Using public magnification maps, we reconstruct the [CII] emission in the source plane to reveal a stable, rotation-dominated disk with $V/\sigma=5.3$, which is $>2\times$  higher than predicted from simulations for similarly high-redshift, low-mass galaxies. 
In the source plane, the [CII] disk has a half-light radius of 1.8$\,$kpc and, along with the dust, is spatially offset from the peak of the stellar light by 1.4$\,$kpc. 
Az9 is not deficient in [CII]; $L_{\rm{[CII]}}/L_{\rm{IR}}=0.0027$ consistent with local and high redshift normal star forming galaxies. 
While dust-obscured star formation is expected to dominate in higher mass galaxies, such a large reservoir of dust and gas in a lower mass disk galaxy 1.4$\,$Gyr after the Big Bang challenges our picture of early galaxy evolution. 
Furthermore, the prevalence of such low-mass dusty galaxies has important implications for the selection of the highest redshift dropout galaxies with JWST. 
As one of the lowest stellar mass galaxies at $z>4$ to be detected in dust continuum and [CII], Az9 is an excellent laboratory in which to study early dust enrichment in the interstellar medium.

\end{abstract}


\section{Introduction} \label{sec:intro}

Our census of the dust content in galaxies at $z>3$ is incomplete due to current observational limitations. While at $z<3$ the dust-obscured star formation is $6\times$ higher than the unobscured star formation \citep{Madau2014}, the expectation is that, at higher redshifts, the obscured star formation will become less dominant. Fundamentally, this is because we expect less dust in the early Universe as it takes time for generations of stars to produce and distribute dust \citep{Popping2017}. In addition, the mass-metallicity relation implies that lower mass galaxies should have less dust \citep[e.g.][]{Remy-Ruyer2015}, and it is observed that the fraction of obscured star formation decreases with decreasing stellar mass at $z\sim0$--2.5 \citep{Whitaker2017}. While the detection of dust continuum at higher redshifts and in lower mass galaxies provides crucial constraints on the formation of dust and metals \citep[e.g.][]{Laporte2017}, this parameter space remains poorly explored. 

ALMA has the sensitivity to detect dust in normal\footnote{We use ``normal" to refer to galaxies which are typical star-forming galaxies for their epoch; on the star-forming main sequence and/or with stellar masses near the knee of the stellar mass function.} galaxies at $z>4$ \citep{Capak2015,Watson2015,Willott2015,Laporte2017,Bethermin2020,Inami2022}. These studies show mixed results: some sources have significant dust emission while others remain undetected \citep[e.g.][]{Schaerer2015,Bouwens2016}. For UV-selected samples, the dust-obscured star formation only dominates in high mass galaxies \citep{Fudamoto2020,Algera2023}, consistent with the trends at $z=0$--2.5 \citep{Whitaker2017}, although a significant population of dusty low-mass galaxies cannot be ruled out. 

An interesting recent development is the recognition that some fraction of the highest redshift ($z>10$) candidate galaxies selected from JWST surveys might actually be $z\lesssim6$ dusty galaxies \citep{Naidu2022,Zavala2022}. With exceptionally bright optical emission lines, a relatively low-mass dusty galaxy at $z\sim5$ can mimic the observed near-IR colors of a $z>10$ candidate \citep{Naidu2022,McKinney2023}. Our lack of prior information on the ubiquity of both $z>10$ galaxies and low-mass dusty galaxies at $z>4$ limits our ability to correctly identify and separate these populations in JWST surveys. 

In this letter, we present observations of gas and dust in a unique galaxy at $z=4.3$: MACS0717\_Az9 (hereafter Az9) is a multiply-imaged galaxy which clearly deviates from the assumption that dust is unimportant in high-redshift, low-mass galaxies. AzTEC imaging on the Large Millimeter Telescope (LMT) revealed substantial dust-obscured star formation (80\%) for this low-mass main-sequence galaxy \citep{Pope2017}. 
Here, we report [CII] and dust continuum detections with ALMA to measure the spectroscopic redshift, put constraints on the ISM conditions, and describe the kinematics and spatial distribution of gas and dust in this galaxy. We aim to understand the extreme dustiness of Az9 and how it relates to other high-redshift galaxy populations. We assume a standard $\Lambda$CDM cosmology with $\Omega_{M}$ = 0.3, $\Omega_{\Lambda}$ = 0.7, and $H_{0}$=70 km/s/Mpc.

\section{ALMA observations} \label{sec:obs}

The HST-identified multiply-imaged system in the Hubble Frontier Fields (HFF) cluster MACS J0717.5+3745 has 3 components 5.1, 5.2 and 5.3 and a photometric redshift of $z\sim4$--5 \citep{Zitrin2009,Diego2015,Limousin2016}. The two most strongly magnified images of this system, 5.2 and 5.1, were detected with AzTEC \citep{Pope2017}. In this paper, we followup the component of this system with the highest amplification, 5.2, and refer to it as Az9. 

\begin{figure*}[hb]
\includegraphics[trim=100 80 150 80,clip,scale=0.95]{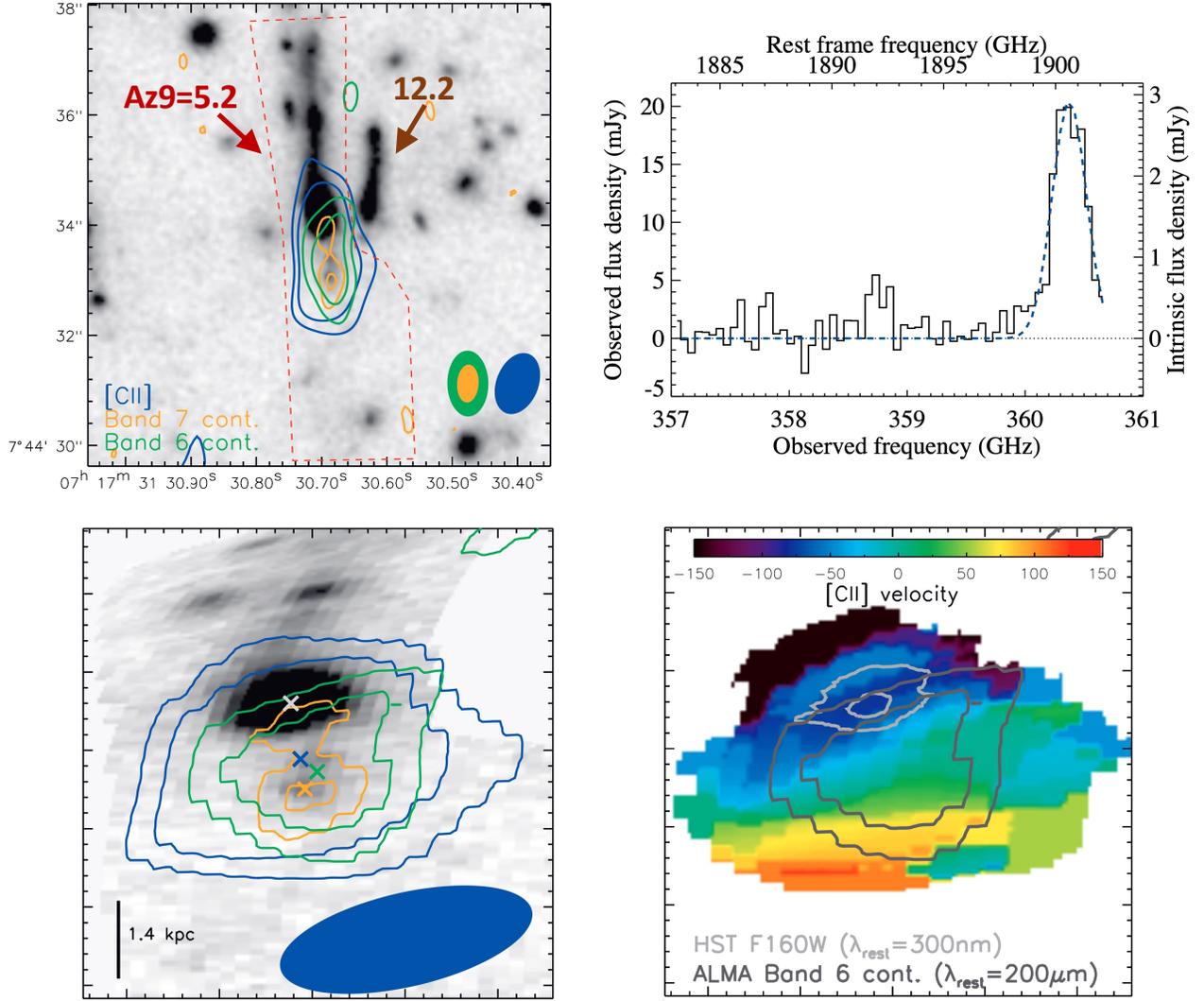}  
\caption{({\it top left}) Image plane showing the HST H-band image with the ALMA [CII] and dust continuum contours (3 and 5$\sigma$). The corresponding ALMA beams are shown in the bottom right. Az9 is the multiple image source 5.2 and we rule out any millimeter emission coming from another nearby source 12.2. The red dashed lines show the region we used for the HST source plane reconstruction. 
({\it top right}) Extracted 1D spectrum (from image plane cube) of Az9 (black) and bestfit Gaussian to the [CII] line (dashed blue line).
({\it bottom left}) Source plane reconstructed images: HST $H$-band ($\lambda_{\rm{rest}}=0.3\,\mu$m) image with the ALMA [CII] and dust continuum contours (3 and 5$\sigma$). Image cutout size is 1.2 arcsec $\times$1.2 arcsec. The colored crosses denote the centroids of each image. The ALMA [CII] reconstructed beam is shown in the bottom right corner.
({\it bottom right}) Source plane reconstructed [CII] moment 1 map with light and dark gray contours showing the HST and Band 6 dust continuum, respectively.  
\label{fig:data}}
\end{figure*}

Az9 was observed in Band 6 continuum in April/May 2018 for 5.6 minutes on-source (2016.1.00293.S, PI: Pope).
Data was reduced with CASA 5.1.2 and cleaned interactively with natural weighting down to $3\sigma$. Continuum emission from Az9 is clearly detected and spatially resolved (top-left panel of Figure \ref{fig:data}). The 1.13$\,$mm flux is extracted using an optimized elliptical aperture and the uncertainty comes from taking the standard deviation of integrated flux measurements in 100 random apertures of the same size offset from the source on the non primary beam corrected image (see Table \ref{tab:source} for image properties). 

A Band 7 spectral sweep was approved to search for [CII] from Az9 (2017.1.00091.S, PI: Pope). We manually designed 9 science blocks to provide uniform sensitivity and cover [CII] from $z\sim4$--5 this range (316.2--372.7$\,$GHz). Only 3/9 science blocks ($a$, $b$, $i$) were observed between June-September 2018, providing 1/3 of the requested spectral coverage. 
The data are reduced using CASA 6.5.0-15 and interactively cleaned using \texttt{tclean} with a robust parameter of $0.5$ at a spectral resolution of $50\,\mathrm{km\,s^{-1}}$. Despite having only 1/3 of the requested bandwidth, a bright line is clearly detected in the $i$ science block cube. We extract the 1D spectrum in the image plane through an optimized aperture and fit it with a Gaussian. The integrated flux of [CII] has a SNR of 24 (Table \ref{tab:source}). The top right panel of Figure \ref{fig:data} shows the spectrum in black and the best-fit Gaussian as the blue dashed line. 

In addition to the spectral cube, we create a band 7 continuum image using the side bands in the $a$ and $b$ science blocks ($\nu=320\,$GHz). The $a$ and $b$ science blocks were observed in a more extended configuration and resulted in a smaller beam. The 0.94$\,$mm continuum is clearly detected and spatially resolved (top-left panel of Figure \ref{fig:data}) in the $a$ and $b$ blocks. The continuum flux is extracted using the same aperture as the Band 6 continuum and we use the same technique described above to calculate the uncertainty on the integrated continuum flux. Data parameters are listed in Table \ref{tab:source}.

\section{Analysis} \label{sec:analysis}

\subsection{SED fitting}

With the latest optical catalogs from HFF-DeepSpace group \citep{Shipley2018}, the upper limits from {Herschel}/SPIRE \citep{Rawle2016} and the measured submm/mm fluxes (Table \ref{tab:source}), we correct the photometry for the known magnification (average of $\mu=7$ over Az9\footnote{https://archive.stsci.edu/prepds/frontier/lensmodels/\#magcal}) and use \texttt{MAGPHYS\_highz} \citep[v2;][]{Battisti2020} to model the full SED. \texttt{MAGPHYS} \citep{daCunha2008,daCunha2015} is designed to self-consistently determine galaxy properties based on an energy-balance approach using rest-frame UV through radio photometry in a Bayesian formalism. In brief, \texttt{MAGPHYS} uses the stellar population models of \cite{Bruzual2003}, assumes a \cite{Chabrier2003} initial mass function, and uses the dust model of \cite{Charlot2000}. We refer readers to the references above for more details. Simulations have shown that \texttt{MAGPHYS} does a good job of recovering physical parameters of isolated galaxies and major mergers except during the near-coalescence phase \citep{Hayward2015}.

\begin{figure*}[hb]
\includegraphics[trim=0 120 0 0,clip,scale=0.7]{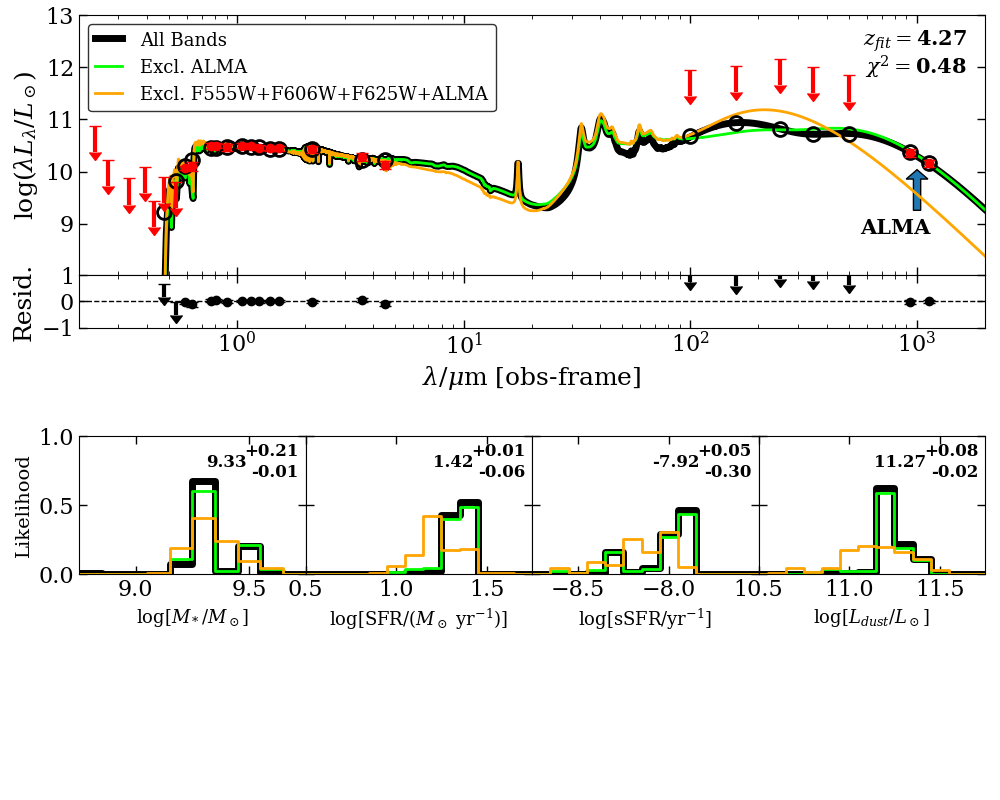}  
\caption{({\it top}) Spectral energy distribution of Az9 (red symbols) with the best-fit model from \texttt{MAGPHYS\_highz} (black open circles and curve). 
Data with S/N$<3$ are treated as 3$\sigma$ upper limits (red down arrows). 
The colored curves show the best-fit model excluding the ALMA data (green) and excluding the ALMA, F555W, F606W, and F625W data (orange). Interestingly, the F555W, F606W, and F625W bands, which cover the range between the Lyman limit and the Lyman-alpha line, appear to place strong constraints on the dust emission.
({\it bottom}) Histograms showing the best-fit values for stellar mass, SFR, sSFR and dust luminosity for the three models shown in the top panel. The median values and uncertainties (from 16th and 84th percentiles) for each property from the best-fit model to all bands (black) are shown in the upper-right of each panel. The derived parameters are consistent for the fits with and without the ALMA data.
\label{fig:sed}}
\end{figure*}

The best-fit SED model is shown in Figure \ref{fig:sed} and the best-fit parameters and their uncertainties are in Table \ref{tab:source}. It is reasonable to question the energy balance assumption in the SED fitting, especially since we find spatial offsets between the optical and infrared light (top left panel of Figure \ref{fig:data}, see discussion in Section \ref{sec:dist}). In order to test this, we re-fit the SED excluding the IR data points. All parameters derived from the SED fitting, including the stellar mass, are completely consistent with the fits that include the IR points (see distributions in the bottom panel of Figure \ref{fig:sed}). Interestingly we find that when the IR bands are excluded, the best-fit SED still lines up perfectly with the ALMA points and predicts the same IR luminosity. This might be surprising but it seems that the bands between the Lyman limit and Lyman-alpha line (F555W, F606W, F625W) are setting strong constraints on the dust. This is shown as the orange curve which is the best-fit excluding the ALMA and F555W, F606W, F625W data points. This appears to be a consequence of the dust attenuation curve parameterization used in MAGPHYS and supports the energy balance argument even though there are offsets in the emission regions.

 Outputs from \texttt{MAGPHYS} include the stellar mass, star formation rate (SFR$_{\rm{SED}}$) and the IR and UV luminosities ($L_{\rm{IR}}$, $L_{\rm{FUV}}$). SFR$_{\rm{SED}}$ is the sum of the stellar mass formed in the last 100$\,$Myr, and can be compared to the sum of the unobscured and obscured SFRs estimated from $L_{\rm{FUV}}$ and $L_{\rm{IR}}$, respectively. 

$L_\mathrm{FUV}$ is calculated by fitting the UV continuum ($1250\text{\AA}\le\lambda\le2600\text{\AA}$) as a power-law with $\beta$ as the UV slope. The UV slope based directly on the photometry is $\beta_\mathrm{phot}=-1.10\pm0.26$.  
We obtain $L_\mathrm{1600}=(3.09\pm0.36)\times10^{10}~L_\odot$ and convert this to FUV based on $L_\mathrm{FUV}\sim0.97\times L_\mathrm{1600}$. 
$L_\mathrm{IR}$(8-1000$\mu$m) is calculated by converting $L_\mathrm{dust}$ from \texttt{MAGPHYS}, which is the integrated dust emission at all wavelengths. 
$\mathrm{SFR_{UV}}$ and $\mathrm{SFR_{IR}}$ are then calculated using the relations found in \cite{Murphy2011}. All values are given in Table \ref{tab:source}. 

The sum of the obscured and unobscured SFRs; $\mathrm{SFR}_\mathrm{IR}$+$\mathrm{SFR}_\mathrm{UV}=30.3~M_\odot/\mathrm{yr}$, is only 15\% larger than $\rm{SFR}_{\rm{SED}}$. This can be attribute to the different assumptions inherent to each method \citep[e.g.][]{Utomo2014}, and suggests the energy balance assumption in \texttt{MAGPHYS} is reasonable.

\subsection{Source plane reconstructions} \label{sec:source_plane}

We use the public lensing models provided through the HST Frontier Fields program\footnote{https://archive.stsci.edu/prepds/frontier/lensmodels/} to perform source plane reconstructions. Specifically, we use the CATS non-cored model \citep{Limousin2016} which predicts a redshift of $4.1\pm0.2$ for Az9, consistent with our new spectroscopic redshift from [CII]. In HST, Az9 is extended N-S in the image plane over about 3 arcsec, and the dust and [CII] extends another 2 arcsec to the south. The range in the magnification across the source (in HST and ALMA) is $\mu=6$--8.5, and the average magnification is $\mu=7\pm1$. The region used for the reconstruction of the HST image is shown with the red dashed lines in the top left panel of Figure \ref{fig:data}; it includes all the clumpy faint emission that may be associated with Az9 and excludes another known multiply imaged source 12.2.

Source-plane reconstructions at a redshift of $z=4.27$ are performed with Lenstool \citep{Kneib1996,Jullo2007,Jullo2009} using the publicly-available CATS parameter files, which explicitly contain the optimized set of lens mass profiles (also determined with Lenstool). For each image, pixels in the image plane are oversampled by a factor of 16 to provide greater accuracy, and source-plane pixels are chosen to oversample by a factor of 8 (reflecting the magnification provided by lensing). For spectral cubes, the reconstruction is performed channel-by-channel. 

We generate reconstructed images of [CII], Band 7 continuum, Band 6 continuum, and the {\it HST} H-band image (bottom panels of Figure \ref{fig:data}). In addition, we reconstruct the beam for each ALMA image by placing the observed synthesized beam in the center of the lensed image. 
As magnification varies only modestly across the extent of the source, the source-plane beam likewise does not vary much in the region of interest. 

We measure the continuum fluxes on the reconstructed source plane images and confirm that the integrated fluxes are consistent with the integrated flux in the image plane maps scaled by the average magnification.

\subsection{Kinematic modeling}
\label{sec:kin}

The bottom right panel of Figure \ref{fig:data} shows that the velocity map of [CII] has a clear rainbow pattern consistent with rotation. We quantify this by performing a kinematic analysis on the source plane reconstructed cube using the $\mathrm{^{3D}BAROLO}$ software (BBarolo, \citealt{DiTeodoro2015}), which fits tilted ring models to emission-line data cubes. BBarolo models the geometric properties of the galaxy, which include the geoemtric center ($x_0,y_0$), inclination ($i$) and position angle ($\mathrm{PA}$). The kinematic properties are the maximum velocity ($V_{max}$) and dispersion ($\sigma_0$) of each ring. 

First, we run the \texttt{SEARCH} algorithm in BBarolo which is based on the \texttt{DUCHAMP} algorithm \citep{Whiting2012} to create a noise map. This is equivalent to masking pixels less than three times the RMS in the moment 0 map. We limit the maximum ring radius to be within the \texttt{SEARCH} noise map, approximately $0.^{\prime\prime}35$ along the major kinematic axis. Prior to fitting the data, we assume a thin disk by fixing the scale height of the rings to $z^o = 0.^{\prime\prime}01$, the resulting kinematic parameters are unchanged if we assume a  2x, 2.5x or 10x thicker disk. The initial guesses for $V_{max}$ and $\sigma_0$ are $100\,\mathrm{km\,s^{-1}}$ and $50\,\mathrm{km\,s^{-1}}$ based on initial analysis of the moment 1 and 2 maps, but we note the final derived quantities are not sensitive to these choices. 
We use a two-stage fit to the data by first fitting both the kinematic and geometric parameters using a ring width equal to the spatial resolution: $r_{ring}=0.^{\prime\prime}1$ along major kinematic axis with reconstructed PSF FWHM$=0.^{\prime\prime}24$. This ensures a good fit to the geometric center, inclination, and position angle (PA) given the spatial resolution of our reconstructed image \citep{YungChau2022}. The best-fit inclination angle and PA are 46.6 degrees and 189.2 degrees respectively. We take the best-fit parameters from this fit as inputs into a subsequent fit where we only allow the maximum velocity and dispersion to vary and decrease the ring width by a factor of two to better sample the rotation curve. Throughout these fits, we approximate the PSF as a 2D Guassian with major/minor FWHM and rotation angle derived from a fit to the reconstructed source-plane PSF, which is then convolved with the rotation model prior to calculating residuals. As noted in Section \ref{sec:source_plane}, the source-plane PSF varies little over the extent of the line emission because of modest change in magnification.

\begin{figure*}[ht!]
\plotone{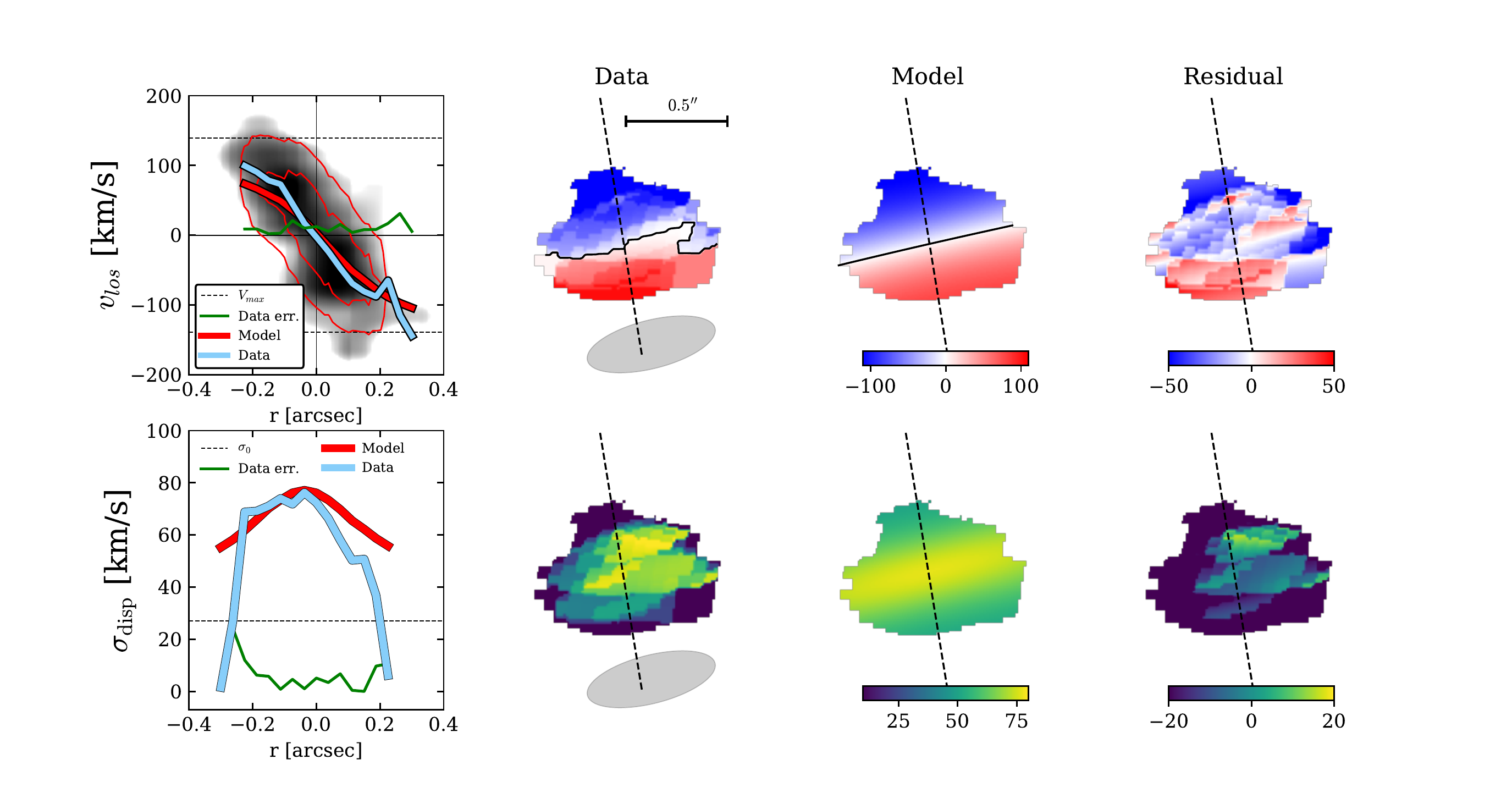}
\caption{Kinematics of [CII] in Az9 in the source-plane image: the top panels show the line of sight rotational velocity, and the bottom panels show the velocity dispersion. 
In the top left inset we plot both the model and data velocities extracted along the major axis over the empirical P-V diagram (black histogram). Contours from the model P-V are shown in red, and uncertainties on the empirical P-V along the major axis are shown with a green line. The bottom left panel shows the velocity dispersion extracted along the major axis including data/model/uncertainties following the top left panel. Moment maps from the reconstructed cube are shown in the data column, followed by maps from the tilted ring model and the residual between the two. The global fit strongly favors a rotating disk with a maximum rotational velocity of $139\,\mathrm{km\,s^{-1}}$ and dispersion of $26\,\mathrm{km\,s^{-1}}$ and reproduces the line-of-sight velocity map along the major axis remarkably well (\textit{top left panel}). The reconstructed beam is shown in grey.  
\label{fig:kin}}
\end{figure*}

The best fit model is shown in Figure \ref{fig:kin} and the derived kinematic parameters are listed in Table \ref{tab:source}. The half-light radius ($r_{1/2}$) is calculated from the azimuthally averaged radial profile of the moment 0 model with errors propagated from the data and the fit. V$_{max}$ is the average rotational velocity of all rings with $r>0.05^{\prime\prime}$ because the inclination-corrected rotation curve is consistent with being flat at these radii. We do not expect high gas dispersion away from the core of the galaxy if it is disk-dominated, and indeed the gas dispersion in the most extended rings are zero within their respective uncertainties. Therefore, we take the average dispersion over all rings with non-zero dispersion (all at $r<r_{1/2}=0.26^{\prime\prime}$) to be $\sigma_0$. We estimate $1\sigma$ uncertainties on derived quantities following the Monte Carlo scheme built into BBarolo, which randomly samples models around the minimized residual (see \citealt{DiTeodoro2015} for further details). For parameters like $V_{max}$ and $\sigma_0$ we propagate the uncertainty per ring into the final averaged quantity's error. 

The moment 2 residuals in Fig.~\ref{fig:kin} show a peak in the northern half of the galaxy, possibly arising from outflowing gas not associated with disk rotation. We re-fit the kinematic parameters using only the southern/red-shifted portion of the galaxy, and find the region of high dispersion persists. The dispersion residual is a factor of two lower than when fitting the entire moment map, and within measurement uncertainty. Given that the residuals improve when masking this region, the dispersion peak in the moment 2 map in Fig.~\ref{fig:kin} is most likely associated with measurement uncertainty. We conclude that our data does not have sufficient signal-to-noise to support the presence of outflowing gas.

\section{Results} \label{sec:results}

\subsection{Spectroscopic redshift}

The bright [CII] detection provides a spectroscopic redshift of 4.2738 for Az9 (top right panel of Figure \ref{fig:data}). Given the brightness of the line and the redshift priors, this line can only be identified as [CII]. The [CII] spectroscopic redshift is consistent with the previous redshift estimates from optical photometry \citep{Pope2017} and the lens modeling \citep{Diego2015,Limousin2016}. Previous optical spectroscopic efforts failed to identify a redshift for this UV-selected galaxy. [CII] may therefore be one of the best ways to identify redshifts in lower mass dusty galaxies. 

As discussed in \cite{Pope2017}, the AzTEC beam (8.5 arcsec) covered both source 5.2 (also known as Az9) and another multiply-imaged galaxy, source 12.2, which is at $z_{\rm{spec}}=1.71$. With the high spatial resolution ALMA detections, we can now definitively rule out any millimeter emission coming from 12.2 (top left panel of Figure \ref{fig:data}. Furthermore, the line we attribute to [CII] cannot be from 12.2 since there are no known lines at the corresponding rest frequency. 

\cite{Treu2015} report a probable HST grism redshift of 0.928 for 5.1, the second brightest image of this system. This redshift is inconsistent with the line we detect in the ALMA spectrum for 5.2 and we conclude that the grism redshift for 5.1 is incorrect.

\subsection{Integrated properties}

\begin{figure*}[hb]
\includegraphics[trim=20 70 40 200,clip,scale=0.45]{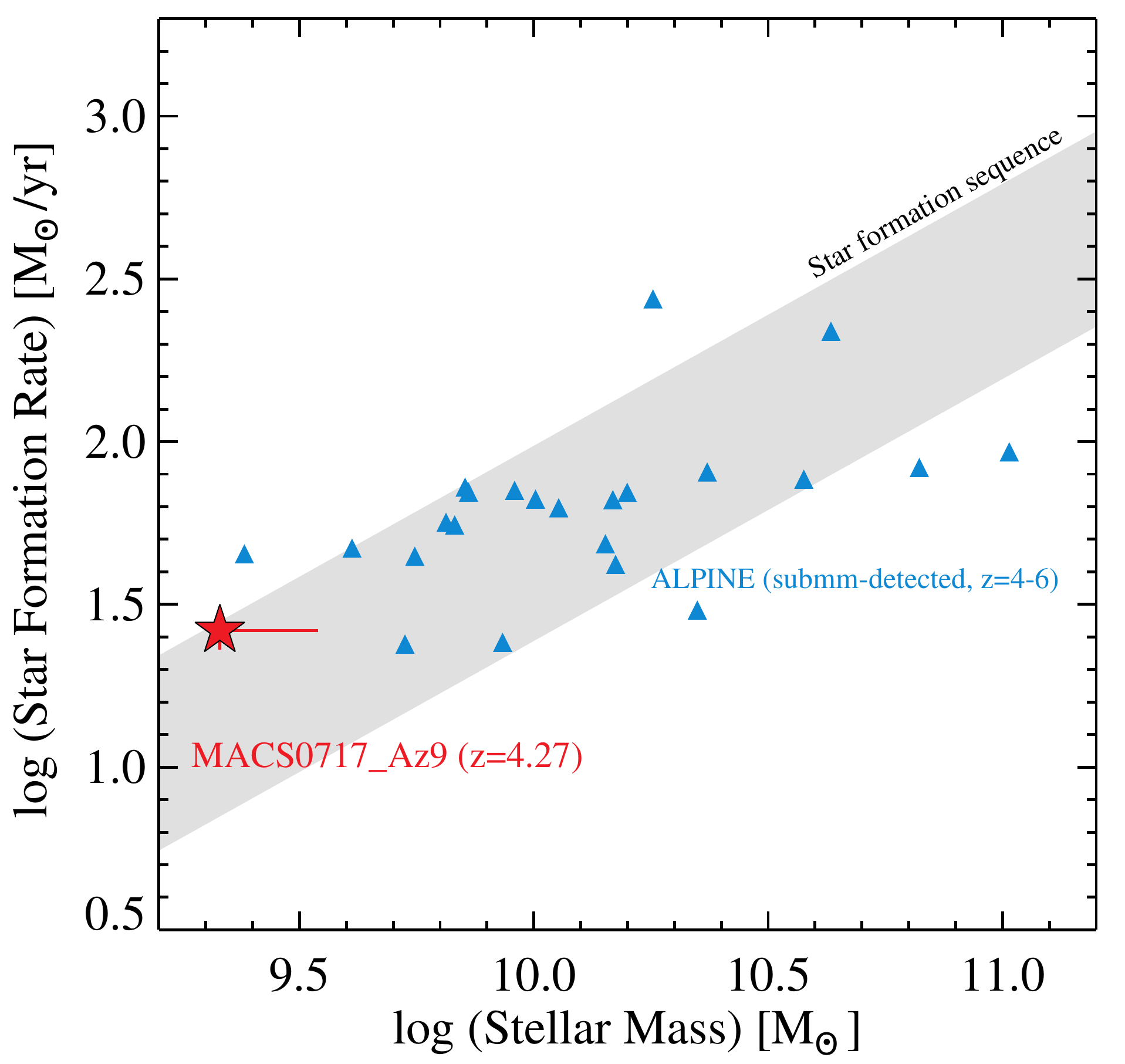}  
\includegraphics[trim=0 70 40 200,clip,scale=0.45]{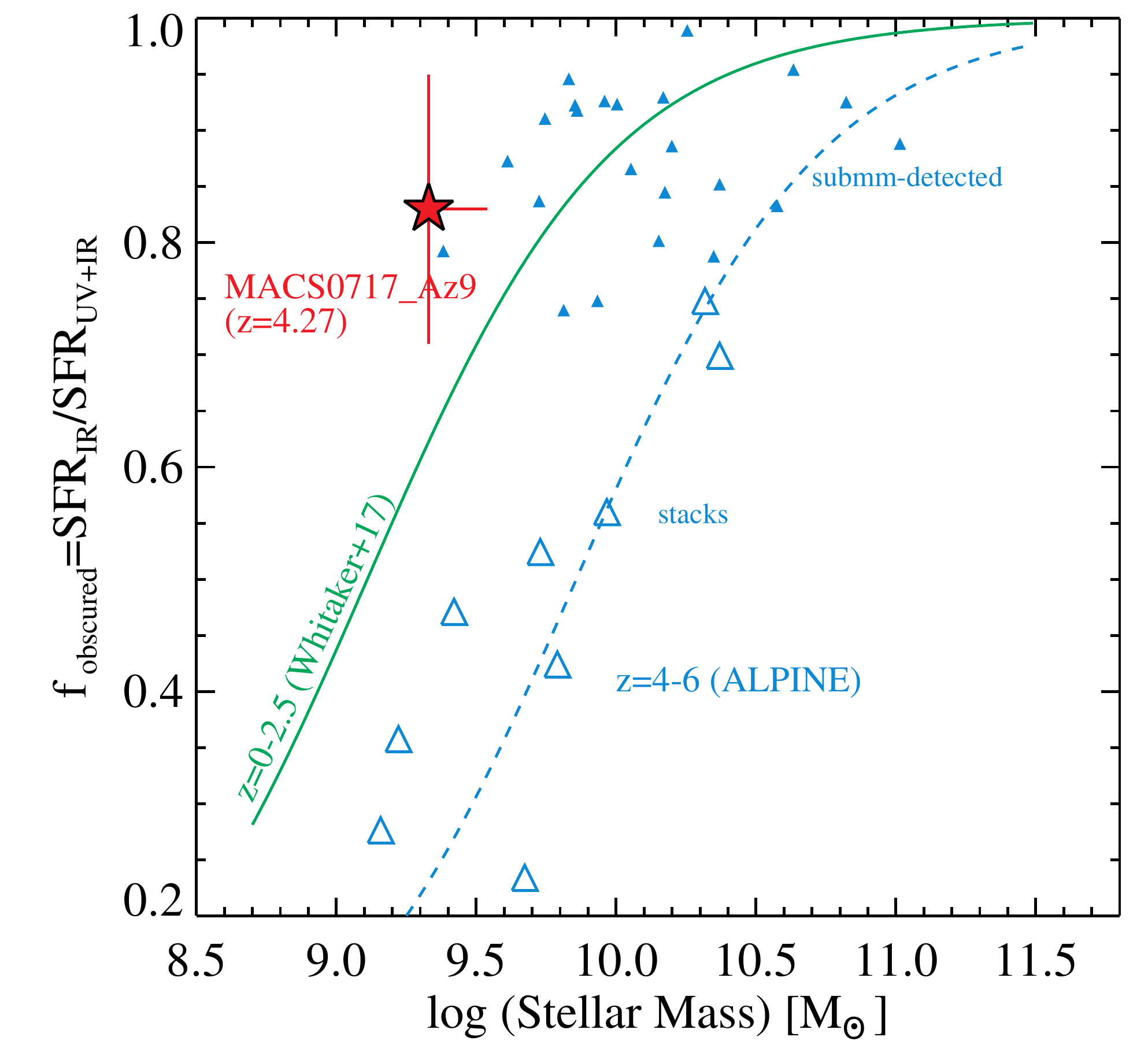}  
\caption{({\it left}) Total star formation rate as a function of stellar mass for galaxies at $z\sim4$--5. Gray shaded region shows $\pm1\sigma$ from the galaxy main sequence at $z=4.5$ from \cite{Speagle2014}. Az9 is shown as the red star in both panels. The blue filled triangles are individual galaxies from ALPINE that are detected in dust continuum. 
({\it right}) Fraction of star formation that is obscured by dust ($f_{\rm{obscured}}$) as a function of the stellar mass. Open triangles show the stacked measurements for the ALPINE sample which are lower in $f_{\rm{obscured}}$ than the subset of sources detected in dust continuum (filled triangles). The green curve is the best-fit relation for $z=0$--2.5 from \cite{Whitaker2017} while the blue dashed curve is the same function scaled down \citep[$a=6\times10^{9}$ and $b=-2.284$ following Equation 1,][]{Whitaker2017} to fit the ALPINE stacked measurements.  
\label{fig:MS}}
\end{figure*}

From the \texttt{MAGPHYS} SED model (Figure \ref{fig:sed}), the intrinsic best-fit stellar mass is log($M_{star})=9.33^{+0.21}_{-0.01}$, sitting below the estimated knee in the stellar mass function at $z\sim4$ \citep{Muzzin2013} and probing an unexplored region of stellar mass at this epoch. The stellar mass determined in this work, with the spectroscopic redshift and updated photometry, is slightly lower, although consistent within the uncertainties, than the stellar mass calculated using a different SED fitting code in \cite{Pope2017}. 
With a $\rm{sSFR}=12\,\rm{Gyr}^{-1}$, Az9 is within the scatter of the star formation main sequence for its redshift (left panel Figure \ref{fig:MS}). 

The width of the [CII] line ($282\pm18\,$km/s) is consistent with the average found for UV-selected galaxies at $z=4$--6 from ALPINE \citep{Bethermin2020} and much narrower than the millimeter-selected galaxies from SPT \citep{Gullberg2015}. We measure a total intrinsic [CII] luminosity of $4.63\pm0.18\times10^{8}\,L_{\odot}$ and $L_{\rm{[CII]}}/L_{\rm{IR}}=0.0027$. Az9 is not deficient in [CII] like similarly dusty high-redshift galaxies \citep[e.g.][]{Gullberg2015}, and has $L_{\rm{[CII]}}/L_{\rm{IR}}$ consistent with measurements of star forming regions in nearby star forming galaxies and the high-redshift ALPINE sample \citep{Smith2017,Schaerer2020}. 

For its redshift, Az9 is a typical galaxy in terms of $L_{\rm{[CII]}}$, SFR and stellar mass. However, unlike most low-mass galaxies, Az9 is very dust obscured. The right panel of Figure \ref{fig:MS} shows the fraction of the SFR that is obscured by dust as a function of stellar mass. Az9 is above the best fit relation for $z=0$--2.5 (green), and its $f_{\rm{obscured}}$ is $>4\times$ higher than the relation fit to the full ALPINE sample at $z=4$--6 \citep[blue dashed curve fit to open blue triangles, data from][]{Bethermin2020}. 
Put another way, the stellar mass of Az9 would need to be an order of magnitude larger to sit on the $f_{\rm{obscured}}$ relation for the ALPINE sample. Additional systematic uncertainties from SED fitting and the lens modeling are insufficient to underestimated the stellar mass so severely; Az9 would remain overly dusty for its stellar mass.

\subsection{Spatial distribution of gas, dust and stars}
\label{sec:dist}
Az9 is clearly detected in the Band 6 and Band 7 continuum images (left panels of Figure \ref{fig:data}). In both cases, the emission is resolved and we measure the fluxes through an optimal aperture in the image plane. The observed fluxes are presented in Table \ref{tab:source}. 

In both the image and source plane, the dust continuum emission is offset from the HST H-band emission suggesting the dust-obscured activity and stellar emission are coming from different regions of the galaxy. The HST emission is probing a rest-frame wavelength of $0.3\mu$m, which is highly susceptible to dust attenuation and may give rise to spatial offsets. While there is some faint HST emission under the [CII] and dust continuum centroids and a bit of clumpiness to the north, the bulk of the HST emission is centered on a region offset by 0.2 arcsec (1.4$\,$kpc) from the center of the dust continuum contours. This offset is consistent with the average offset found between UV and IR emission in galaxies in ALPINE \citep{Fujimoto2020}. Interestingly, the unobscured stellar light is coincident with the blueshifted [CII] emission (bottom right panel of Figure \ref{fig:data}). While it is possible an outflow is clearing the dust on the blueshifted side of the disk, we see no evidence of this in our kinematic analysis (Section \ref{sec:kin}). 

From the kinematic modeling, we obtain a [CII] half-light radius of 0.26 arcsec, which corresponds to 1.8$\,$kpc at $z=4.274$. This radius is consistent with the range of [CII] sizes measured for lower-mass galaxies in ALPINE \citep{Fujimoto2020}.

\subsection{A stable rotating gas disk}
\label{sec:rotating}

Figure \ref{fig:kin} shows the results of our kinematic modeling. Az9 shows a smooth velocity gradient across the galaxy, defining the kinematic axis, and a centrally peaked velocity dispersion distribution. We calculate $V/\sigma=5.3\pm3.6$, which is consistent with a stable, dynamically-cold, rotating disk. 
We compare our measured $V/\sigma$ with values found in simulations and in other observations but note the caveat that different definitions and ways of measuring $V$ and $\sigma$ may lead to different values.
Az9 has a $V/\sigma$ that is $>2\times$ higher than the $V/\sigma$ predicted for similarly low-mass galaxies at $z=4.3$ from the TNG50 simulation, driven primarily by the higher velocity dispersion predicted in the simulation \citep{Pillepich2019}. With clear rotational disk kinematics, Az9 does not show any indications of being a merger. 

\cite{Rizzo2022} test the robustness of $V/\sigma$ at classifying disks and mergers using mock [CII] observations and find that there are some cases where a merger can look like a disk in $V/\sigma$; however, the mergers are always picked out with multiple-peaked emission profiles. For our kinematic analysis, we have $\sim1.5\times$ resolution elements over the major kinematic axis; in this regime disk galaxies are always classified correctly but $\sim50\%$ of mergers are mis-classified as disks \citep{Rizzo2022}. While we cannot rule out a merger based on $V/\sigma$ alone, the [CII] profile of Az9 is nicely single peaked (Figure \ref{fig:data}) in favor of the disk interpretation.

There are only handfuls of rotating disks observed at $z>4$ and even fewer at low stellar masses. 
Observations show that the fraction of rotation-dominated (disk) star forming galaxies with stellar mass $<10^{10}M_{\odot}$ at $z=3$ is $<40\%$ \citep{ForsterSchreiber2020}, with the disk fraction increasing with stellar mass \citep{Tiley2021}. Of the 29 [CII]-detected galaxies in ALPINE with detailed kinematic modeling with BBarolo, only 6 (21\%) were classified as rotators with stellar masses $\sim10^{10}M_{\odot}$ \citep{Jones2021}. \cite{Rizzo2020} present a remarkable rotator at $z=4.2$, surprisingly unaffected by a nearby companion \citep{Peng2023}, but with a stellar mass $6\times$ higher than Az9 \citep[see also][]{RomanOliveira2023}. 
\cite{Isobe2022} find that local galaxies with low masses ($<10^{9}M_{\odot}$) are all observed to have $V/\sigma<1$. Az9 therefore appears to be an outlier compared to existing observations with higher $V/\sigma$ for its stellar mass and redshift. 

In addition to showing clear rotation, the low velocity dispersion of Az9 ($26\,$km/s) suggests it is stable. Dusty star-forming galaxies at $z\sim2$ from the KAOSS survey have average rotational velocities and velocity dispersions from ionized gas of 190$\,$km/s and 90$\,$km/s, respectively \citep{Birkin2023}. Even though these dusty galaxies are technically rotation dominated, their high dispersion suggests turbulent rotating disks. 
Even accounting for the fact that the ionized gas dispersion is observed to be $\sim10$--15$\,$km/s higher than the molecular/atomic gas dispersion in galaxies out to $z\sim2.6$ \citep{Ubler2019}, Az9 still has a much lower velocity dispersion. This may be expected for its lower mass, but perhaps unexpected given its high dust content. It is unclear what role dust plays in disk turbulence and measuring the kinematics of gas in multiple phases in higher redshift and lower mass galaxies will help address this question.

From the rotational velocity and the radius, we calculate a dynamical mass at $2\times r_{1/2}$ of $M_{\rm{dyn}}=1.6\times10^{10}M_{\odot}$ which gives $M_{\ast}/M_{\rm{dyn}}\sim0.1$. Measurements of the molecular gas mass are needed to complete the census of baryonic mass in Az9 and constrain the dark matter fraction.

\section{Discussion} \label{sec:discussion}

We report the discovery of a dynamically cold, rotating disk in an unusually dusty, low-mass galaxy (known as Az9), which sits on the star forming main sequence at $z=4.3$. 
While its low stellar mass would suggest a lower metallicity at this early epoch (1.4$\,$Gyr after the Big Bang), the large amounts of dust on the contrary (implied by the dust-obscured SFR) predict that significant metals must already be in place. 
For resolved regions of nearby galaxies and integrated emission from high-redshift galaxies, $L_{\rm{[CII]}}/L_{\rm{IR}}$ decreases as a function of the star formation surface density ($\Sigma_{\rm{SFR}}$), and for a given $\Sigma_{\rm{SFR}}$, higher values of $L_{\rm{[CII]}}/L_{\rm{IR}}$ have lower gas-phase metallicity \citep{Smith2017}. Az9 sits on the top edge of this trend, with higher $L_{\rm{[CII]}}/L_{\rm{IR}}$ for its $\Sigma_{\rm{SFR}}$ (Table \ref{tab:source}) which suggests a lower metallicity. The observed amounts of [CII] and dust emission appear to have different implications for the metallicity of Az9. Future ALMA observations of additional FIR fine-structure lines such as [NII] and [OIII] can be used to measure the metallicity of Az9 and put additional constraints on early dust production. 

Another piece of the puzzle for Az9 is the stable, ordered kinematics for such a low-mass, high-redshift galaxy. While simulations and models show that galaxies at $z>4$ are expected to be dynamically hotter and more turbulent than lower redshift galaxies, Az9 presents a counterexample with clear evidence that even low-mass galaxies can be stable against the harsher conditions in the early universe. Linking the gas and dust fractions to the kinematics is an important step in understanding the role of turbulence in how galaxies evolve.

While Az9 is an outlier compared to existing UV-selected populations, it remains to be seen whether there is a larger population of heavily-obscured, low-mass galaxies. These galaxies will not have been selected in the UV in ALPINE and REBELS due to their low mass and extreme dustiness. 
Taking census of dusty, low-mass galaxies in the rest-frame UV/optical is only now possible with increased wavelength and sensitivity of JWST. Several early JWST papers have confirmed that previous HST-dark galaxies are dusty disk galaxies \citep{Nelson2022,Barrufet2022}. \cite{Barrufet2022} present a handful of galaxies similar to Az9 in terms of redshift, stellar mass and star formation rate. Other JWST studies have suggested that the very highest redshift galaxy candidates might actually be lower redshift dusty galaxies \citep{Naidu2022} and (sub)millimeter observations are crucial for confirming these results \citep{Zavala2022}. 
Deep surveys with JWST coupled with millimeter observations, such as upcoming surveys with TolTEC\footnote{http://toltec.astro.umass.edu} on the LMT, can show the ubiquity of this dusty galaxy population at lower stellar masses and provide a more reliable selection of the highest redshift galaxies.

\clearpage
\begin{table}
	\centering
	\caption{MACS0717\_Az9 source properties: observed (obs) and intrinsic (int). 
 Intrinsic values are calculated using the CATS 4.1 lensing model \citep{Limousin2016} which has an average magnification of 7 over Az9.}
	\label{tab:source}
	\begin{tabular}{llll} 
		\hline
Band  & Parameter  & Measurement & Units \\
			\hline
6 cont. & $\nu_{\rm{obs}}$ & 265 & GHz \\  
    & beam & $1.20\times0.75$ & arcsec\\
   & rms & 0.085 & mJy/beam \\
   & $S_{\rm{obs}}$ & $0.85\pm0.15$ & mJy \\ 
   & $S_{\rm{int}}$ & $0.121\pm0.021$ & mJy \\

   \hline
7 cont.    & $\nu_{\rm{obs}}$ & 320  & GHz \\ 
& beam & $0.69\times0.39$ & arcsec \\
   & rms & 0.11  & mJy/beam \\
   & $S_{\rm{obs}}$ & $1.10\pm0.18$ & mJy \\
   & $S_{\rm{int}}$ & $0.157\pm0.026$ & mJy \\
   \hline
 7 [CII] & $\nu_{\rm{obs,[CII]}}$ & $360.375\pm0.009$ & GHz \\
   &   beam  & $1.11\times0.76$ & arcsec \\ 
   & $z_{[CII]}$ & $4.2738\pm0.0001$ & \\
   & $V_{\rm{FWHM}}$ & $282\pm18$ & km/s \\
   & Sdv & $5.53\pm0.23$ & Jy km/s\\  
   & $L_{\rm{obs,[CII]}}$ & $3.24\pm0.13$ & $10^{9}L_{\odot}$ \\
   & $L_{\rm{int,[CII]}}$ & $4.63\pm0.18$  & $10^{8}L_{\odot}$ \\
   & $V_{max}$ & $139\pm22$  & km/s \\
   & $\sigma_{0}$ & $26\pm17$  & km/s \\
   & $r_{1/2}$  & $0.26\pm0.07$ & arcsec \\ 
   & $M_{\rm{dyn}}$  & $1.6\pm0.5$ & $10^{10}M_{\odot}$ \\ 
 \hline	
  All  & $L_{\rm{IR,int}}$  & $1.7^{+0.3}_{-0.1}$ & $10^{11}L_{\odot}$ \\
  All  & $M_{\ast,\rm{int}}$  & $2.14^{+1.04}_{-0.05}$ & $10^{9}M_{\odot}$ \\
  All  &  $L_{\rm{[CII]}}/L_{\rm{IR}}$ & 0.0027\\
 \hline	
   All  &  SFR$_{\rm{SED}}$ & $26.3^{+0.6}_{-3.6}$ & $M_{\odot}/\rm{yr}$\\
   All  &  SFR$_{\rm{IR}}$ & $25.2^{+4.4}_{-1.5}$ & $M_{\odot}/\rm{yr}$\\
   All  &  SFR$_{\rm{UV}}$ & $5.1\pm0.6$ &  $M_{\odot}/\rm{yr}$\\
   All  &  $\Sigma_{\rm{SFR_{IR}}}$ & $2.5\pm1.0$ &  $M_{\odot}/\rm{yr}/\rm{kpc^2}$\\
   All  &  $f_{\rm{obscured}}$ & $0.83\pm0.12$ &  \\
 \hline	
	\end{tabular}\\
\end{table}

\begin{acknowledgments}
We thank the referee for their thoughtful and constructive comments which improved the quality of this paper. 
AP thanks Kevin Harrington for insightful conversations, Marceau Limousin for advice with the lensing model and Richard Simon for help optimizing the ALMA science blocks. AM thanks support from Consejo Nacional de Ciencia y Tecnolog\'ia (CONACYT) project A1-S-45680. 
This paper makes use of the following ALMA data: ADS/JAO.ALMA\#2016.1.00293.S, ADS/JAO.ALMA\#2017.1.00091.S. ALMA is a partnership of ESO (representing its member states), NSF (USA) and NINS (Japan), together with NRC (Canada), MOST and ASIAA (Taiwan), and KASI (Republic of Korea), in cooperation with the Republic of Chile. The Joint ALMA Observatory is operated by ESO, AUI/NRAO and NAOJ. The National Radio Astronomy Observatory is a facility of the National Science Foundation operated under cooperative agreement by Associated Universities, Inc. This work utilizes gravitational lensing models produced by PIs Bradac, Natarajan \& Kneib (CATS), Merten \& Zitrin, Sharon, Williams, Keeton, Bernstein and Diego, and the GLAFIC group. This lens modeling was partially funded by the HST Frontier Fields program conducted by STScI. STScI is operated by the Association of Universities for Research in Astronomy, Inc. under NASA contract NAS 5-26555. The lens models were obtained from the Mikulski Archive for Space Telescopes (MAST).

\end{acknowledgments}

\bibliography{myrefs}{}
\bibliographystyle{aasjournal}

\end{document}